# Opto-mechanical expulsion of individual micro-particles by laser-induced shockwave in air

M. C. SCHROEDER[1], U. ANDRAL[1], AND J.-P. WOLF[1,a)]

[1]*Groupe de Physique Appliquée, Université de Genève, 1211 Geneva 4, Switzerland*

[a)]Author to whom correspondence should be addressed: jean-pierre.wolf@unige.ch

**Abstract:** It was recently demonstrated that laser filamentation was able to generate an optically transparent channel through cloud and fog for free space optical communications applications. However, no quantitative measurement of the interaction between the laser-induced shockwave and the aerosol particles was carried out so far, leaving the precise nature of the clearing mechanism up for discussion. A critical question was the maximum distance at which the filament could still act on the aerosol particle. Distances widely exceeding the filament diameter and its energy reservoir exclude other potential clearing effects like shattering or explosion by direct exposure to the laser. Here, we quantify the force exerted by the shockwave on a single aerosol microparticle. The force is measured by observing the ejection and displacement of the particle when trapped in an optical tweezer. We demonstrate that even for distances ranging from 1.5 mm to 5.5 mm away from the filament, thus widely exceeding the filamentary region, an acoustic force of 500 pN to 8 nN (depending on the initial laser power) acts on the aerosol particle and expels it away from the optical trap.

Free space optical communications (FSO) are attractive alternatives to their traditional radiofrequency (RF) counterparts, as they potentially allow 1000 times faster bitrates[1–4] and quantum cryptographic security[5–9]. A severe drawback is, however, higher sensitivity to weather conditions, and in particular to scattering from clouds and fog[10–13]. The mitigation of this limitation currently relies on the multiplication of ground transmitter/receiver stations (« redundancy strategy »), which significantly increases cost and complexity, or dual band operation (RF and FSO), which reduces the appeal of the technology. Cloud coverage is present over Europe 57% of the time on a yearly average[14]. A promising approach was recently proposed, which relies on filamentation of high intensity and high average power lasers[15–18]. The non-linear propagation of high intensity lasers is characterized by a self-guiding behavior, called filamentation, where Kerr self-focusing is compensated by higher order nonlinear effects, like plasma formation. Plasma relaxation then induces a fast (typically 10 ns) heating of the air, up to 1500 K[19], which in turns produces a shock wave[20,21]. This shock wave has been proposed as an efficient opto-mechanical mechanism to expulse water droplets from the laser path and open a clear channel in fogs on a meter length scale[17,22], which at sufficient repetition rates (i.e. kHz), prevents diffusion of new droplets into the channel. This permanently open channel thus allows unperturbed communication in both directions.

An alternative to plasma induced heating was recently proposed by Schroeder et al[23]: the coherent control of the molecular rotation of nitrogen molecules. Control was obtained by Raman exciting $N_2$ with a train of femtosecond pulses with an adjustable inter-pulse period[24]. Clearing was efficiently produced when the interval in the comb was resonant with the rotational period (8.36 ps) of the molecules, while it was inefficient when the interval was off-resonance.

All these previous measurements were carried out on droplets ensembles, generated in a cloud chamber. In order to better understand the opto-mechanical clearing process and quantify its efficiency, the investigation of the interaction of a single filament on a single microparticle at a defined distance is required.

Spontaneously, the use of an electro-dynamic Paul trap for trapping a single microparticle in the vicinity of the filament comes to mind. Successfully used to investigate droplet explosion and ice particle shattering previously[25,26], they proved unreliable in the present case as the microparticle needs to be charged for trapping. The electrostatic interaction between the charged droplet and the charges released by the plasma prevents a precise measurement of the opto-mechanical drag force. Alternatively, a droplet generator synchronized with the laser repetition rate seems attractive. Here the laser-induced shock wave produces a kick on the falling droplet, modifying its trajectory. Such an experiment was carried out in our laboratory, the results of which can be found in the supplementary (Section 1), and recently by Goffin et al.[27]. Although partially convincing, our results were still affected by residual charging of the droplets. The well-known parasitic charging of microdroplets by nozzles[28–30] could be evidenced previously while investigating second and third harmonics generation from microdroplets [31,32]. A perfect neutralization of the droplet stream was impossible in the present investigation and we consequently decided to use optical tweezers for a more precise measurement, inherently immune to charging and cumulative effects.

In this letter, we chose a dual-beam, counter-propagating, open-air trap design, which provides two benefits: (1) easy access to the trapping cell for loading the trap, and (2) scalability of the applied, radial trapping force, by increasing the power of the trapping laser, while maintaining the equilibrium position inside the trap. While a single beam, open-air design would have been easier to realize, the necessity for tuning the trapping force is paramount.

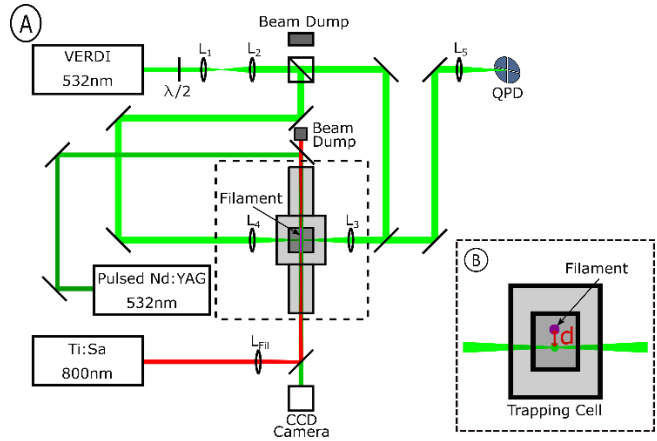

FIG. 1. A) Optical trap used for trapping silica microspheres in the expulsion experiment. Trapping is achieved via two focused beams (532 nm) inside an enclosed cell. Femtosecond pulses are used for the generation of the plasma filament and shockwave inside the cell. A 10 ns pulsed YAG laser provides the background illumination for the shockwave visualization. $L_1$, $L_3$, $L_4$, $L_5$ = 50 mm, $L_2$ = 100 mm and $L_{Fil}$ = 200 mm. B) View from the perspective of the CCD. Notice the distance $d$ between the filament and trapped particle.

The continuous wave Nd:YAG laser (Verdi, $\lambda$ = 532nm, 5W) was split by a polarizing beam splitter and the power in the two paths balanced with a $\lambda/2$ waveplate. Orthogonal polarization reduced interference in the focal region. The alignment of the foci was assured by a removable (5±1) μm pinhole inside the cell.

The trapping cell consists of an outer cube, sealed on all sides by windows, excluding the top, and a smaller inner cell. The inner cell was 3D printed with an electrically conductive filament (*Proto-pasta Conductive PLA*) and grounded, to prevent charge accumulations on the walls. An iris allowed access to the cell from the top for loading the trap with the microparticles. We used silica microspheres of 4.82 μm diameter (*Polyscience*), comparable in size and weight to water droplets in natural fog/clouds [33,34]. Under ideal conditions, microspheres could remain trapped by this optical tweezer for several hours. Using a 20x microscope objective and a CCD (*Thorlabs DCC1545M*) the trapped microspheres could be imaged, as shown in supplementary Fig. S3. Direct imaging was used to verify that only a single sphere, and not a cluster of several, was trapped.

The laser filaments were generated by a Ti:Sapphire (TiSA) laser system ($\lambda = 800$ nm, $< 5$ mJ) that produced pulses at either 1kHz, for the alignment of the filament, or as manually triggered single shots for the particle displacement measurements. The distance $d$ between the trapped particle and the filaments (Fig. 1b) could be varied between 1.5 mm and 5.5 mm. Shorter separation distances lead to immediate microsphere expulsion, due to the strong non-linear scaling of the shockwave amplitude with radial distance and the limited trapping laser power. Larger separations were prohibited by the chamber geometry. The drag force exerted by the filament induced shock wave is opposed by the gradient force of the optical trap, which can be varied by adjusting the trapping laser power output. A direct measurement of the force exerted by the shock wave can thus be performed, provided the optical trap is precisely calibrated beforehand. In particular, the threshold value of the optical gradient force, above which the particle is freed by the filament, gives the measure of the force exerted by the shock wave.

The calculation of the optical trapping force acting on the microsphere was calculated by using Mie-Scattering theory[35], and calibrated by a preceding, independent measurement of the trap stiffness. The stiffness was determined by pushing the trapped microspheres slightly off their equilibrium position with the filament induced shockwave (i.e. at low TiSA laser power), which resulted in an oscillatory behavior. These oscillations in the microsphere position were detected via a quadrant-position detector (QPD), onto which one of the trapping beams was imaged. From the oscillation frequency, the trap stiffness was derived and the optical potential calibrated.

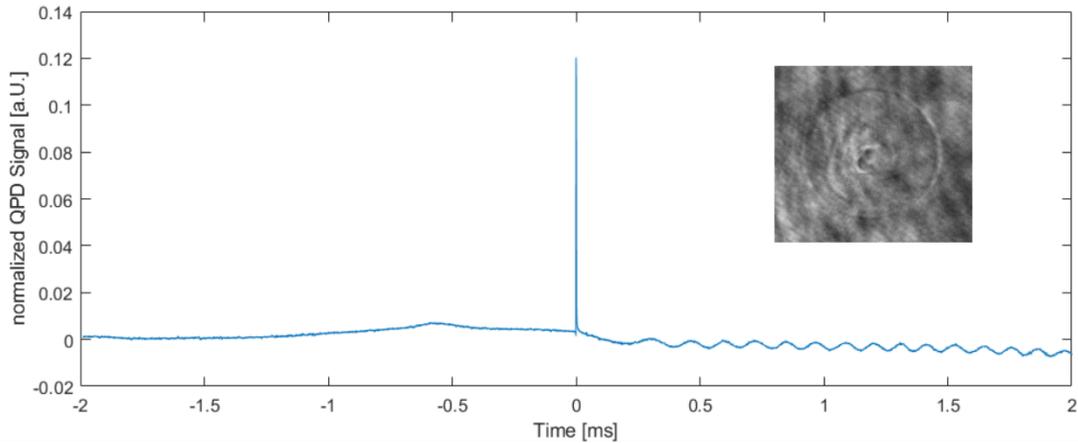

FIG. 2. Laser Induced oscillation for active calibration of dual-beam optical trap. The peak signal indicates the femtosecond laser pulse incident time $t_0$. It results from scattered light hitting the QPD. The oscillations elicited by the opto-mechanical displacement of the trapped microsphere are visible after $t_0$. The oscillation frequency $\omega$ can be gained from the oscillation period $T_0$. The frequency depends on the trap, which changes with the applied trapping laser power. An example image of an induced shockwave is shown in the inset.

The oscillatory behavior mimics that of a damped harmonic oscillator, for small displacements. Figure 2 shows the evolution of the lateral position of a trapped microsphere, recorded by the QPD connected to an oscilloscope. The inset shows the shock wave produced by the filament in the trap, measured by shadowgraphy (size of the image: 1.168mm x 1.101 mm). In the presented case, clear oscillations of the order of ~ 10 kHz are observed. From the resonance frequency, the trap stiffness $k$ is derived, according to $\omega_0 = \sqrt{k/m}$. To determine the resonant radial trap frequency $\omega_0$, the effect of air dampening on the motion of the particle has to be considered as well. The measured frequency $\omega$ is thus given by $\omega = \sqrt{\omega_0^2 - \gamma^2}$, where $\gamma$ is the damping coefficient in air. These measurements were repeated several times at different power levels $P_0$ for the trapping laser, ranging from 100 $mW$ to 800 $mW$. From these measurements, we extracted a calibration curve, shown in Figure 3, which relates the trapping laser power to its corresponding oscillation time, as. $2\pi/T_0 = \omega_0 \propto \sqrt{P_0}$. This calibration curve provides the trap stiffness $k$ for a set of trapping laser power. However, the trap stiffness was used to calibrate our model for distances close to the equilibrium position.

The model, based on the Optical Tweezers Toolbox (OTT) developed by T. A. Nieminen et al.[35], calculates the axial and radial trapping forces acting on a 4.82 μm diameter, silica microsphere in our dual-beam, counter-propagating, open-air optical trap. The theory delivers an analytical solution for simple cases of optical trapping of homogenous and isotropic microspheres under a monochromatic incident light field. This solution is then extended to non-plane wave illumination, to account for the focusing of the incident trapping beams. This extended solution is commonly called the generalized Lorenz-Mie theory (GLMT). In order to calculate the optical forces acting on the trapped microspheres, the toolbox uses the T-matrix formalism in a vector spherical wave basis.

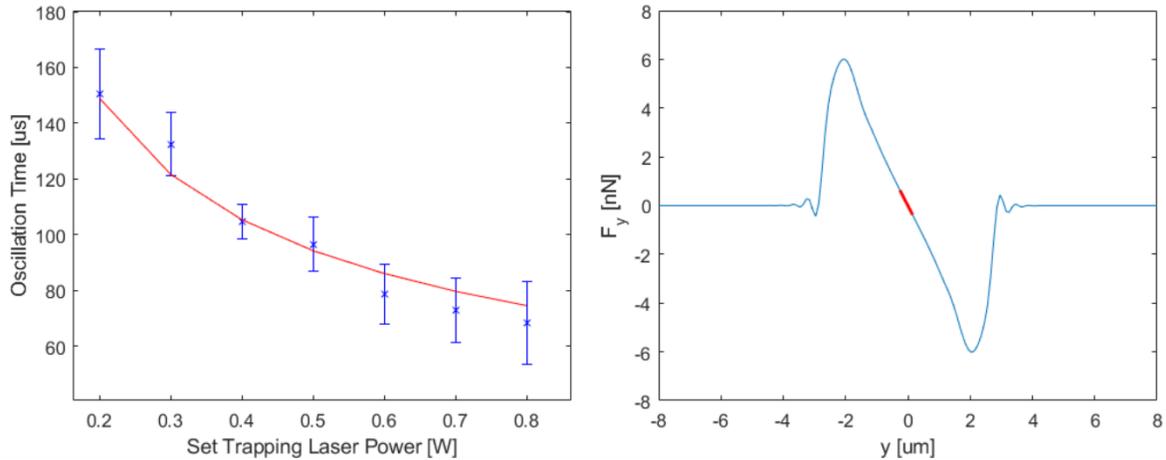

FIG. 3. The left figure shows the calibration curve for the oscillation period $\tau$ based on the trapping laser power. The calibration is fitted according to the $T_0 \propto 1/\sqrt{P_0}$, with the oscillations period $T_0$ in an optical trap and $P_0$ applied laser power. Shown in the right figure is the radial restoring force $F_y$ acting on a trapped microsphere in the dual beam optical tweezer, based on the displacement in the trap orthogonal to the beam axis. In red, the experimental stiffness parameter to which the calculated model has been fitted.

Shown in Figure 3 is the calculated radial trapping force acting on a microsphere. The model is adjusted to overlap with the experimentally determined trap stiffness around the equilibrium position (overlap shown in red). By fitting the model to the measured value for the stiffness, we were able estimate the maximum restoring force, which needs to be overcome for the expulsion of a microsphere. In our measurements we recorded the minimum trapping laser power required to keep the sphere trapped while hit by the filament shock wave. With our model these values were translated into the corresponding restoring force acting on the sphere. This way, we could derive the value of the force exerted by the shockwave onto the silica microspheres. These measurements were carried out at different femtosecond laser pulse powers and separation distances between the trapped sphere and the filament. The measurements give a clearer picture of the scaling between the applied ejection force and the shockwave propagation distance. The results are shown in Figure 4.

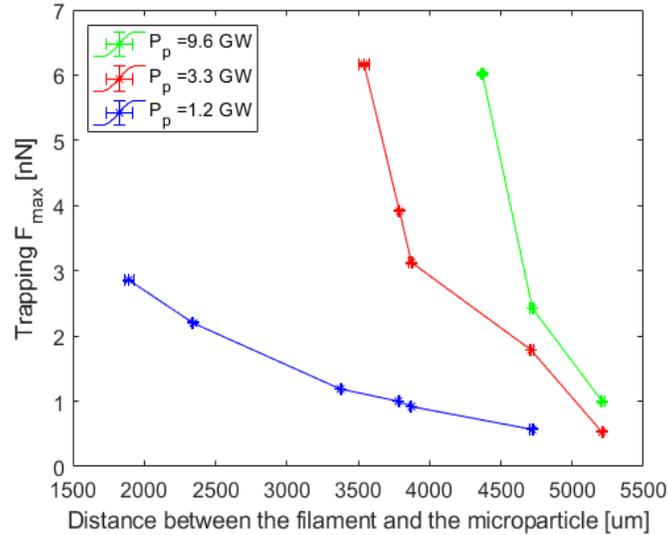

FIG. 4. The curves show the highest radial trapping force able to withstand the effects of the filament induced shockwave, based on the distance between the filament and the trapped microsphere. This force equates the force exerted by the shockwave on the sphere at the given position. The measurements were performed for three incident laser powers.

The propagation distance was determined using the nanosecond Nd:Yag laser illumination beam. The beam was sent in the counter-propagating direction to the femtosecond pulse, as shown in Figure 1a. A digital delay generator was used to synchronize the time between the filament firing and the Nd:Yag probe laser illumination. A CCD camera then recorded the perturbation of the probe laser wavefront by the induced air density perturbation through shadowgraphy. The resulting image shows not only the center of the filamentation region, but also the propagating shockwave. Both can be used for the positioning of the filament and the calculation of the separation distance between it and the trapped sphere. The snapshot shown in Figure 3 shows the cylindrical symmetry of the acoustic wave, as previously observed in similar shadowgraphy experiments [23,36]. The range of the effect was investigated for different particle-filament distances and observed up to 5.5 mm. For very short distances, the depth of the trapping well was never sufficient to counteract the applied acoustic force. For the three investigated incident laser peak powers (1.2 $GW$, 3.3 $GW$, 9.6 $GW$) a strong, non-linear decrease of the force is observed with distance. With the calibration of the optical trap as described in this work (~1.5 $nN/\mu m/W$) and in the case of a perfectly aligned trap, the required ejection force ranges from approximately 3 $nN$ at 1.7 $mm$ to 500 $pN$ at 4.7 $mm$ for the lowest peak power incident laser pulse. At higher peak powers the force could reach up to 6.0 $nN$ at distances larger than 3.5 $mm$. In the case of a sub-optimal alignment of the trap (see supplementary Fig. S5), the required forces are naturally reduced, only between 240 $pN$ and 2.7 $nN$ are needed for expulsion, depending on separation distance and peak power. We observe that the expulsion force strongly increases the peak power of the pulses: for higher peak powers microspheres could be removed from the optical tweezer at further distances and higher set trapping powers.

If we consider a cylindrical shock wave, the amplitude $\Delta P(r)$ of the pressure increase is expected to scale with radius $r$ as $\Delta P(r) \sim P_0\, r^{-q}$ with $0.75 < q < 2$ [37–39] depending on the intensity of the temperature and pressure jumps. Even for very low intensity shock waves, as presented by Bethe et al., the amplitude is characteristically decaying faster than a continuous acoustic wave would with $\Delta P(r) \sim P_0\, r^{-0.5}$. In our case, by fitting the experimental results (Supplementary Figure S6), we find q = 1.6 for the lowest peak power, which fits well the dependency described by Bethe et al. At higher peak powers the results still follow the $\sim r^{-q}$ dependence, but exceed the proposed limits of this model. The development of a shock wave can be much more complex than simple models suggest, as recently shown for the shock wave produced by a few millimeters electric discharge [40,41]. In particular vortices due to viscosity appear, signature of turbulence. The precise shock wave dynamics induced by the filament warrant further study, as they offer a unique opportunity to study these non-linear acoustic phenomena on a table top setup, and without perturbations due to electrodes.

In conclusion, our experiment shows that the shockwave created by the plasma recombination after filamentation exerts a force in the nanonewton range on microspheres surrounding the filament, and this at distances widely exceeding its energy reservoir. This is a significant step towards the understanding of the relevant mechanism of opto-mechanical expulsion of droplets using laser filaments[17]. Our results show that opto-mechanical expulsion is an effective way of creating clear channels in clouds, and that the latter mechanism does not rely on evaporation, or shattering, at least not in a dominant way.

See the supplementary material for the following: Materials and Methods; Section 1, Figure S1, S2 and S3: Discussion of free falling droplet experiment and electrostatic influences; Figure S4: Direct imaging of trapped microspheres; Figure S5: Trapping Force at sub-optimal overlap as a function of restoring force versus filament-microsphere separation distance. Figure S6: Experimental data for of restoring force versus filament-microsphere separation distance fitted with the expected pressure amplitude decay for a cylindrical shock wave.

This research is supported by the Swiss National Foundation for Research under grant number 200021-178926. MCS, UA and JPW thank Prof. Thomas Leisner, Dr. Alexei Kiselev and Dr. Ahmed Abdelmonem for the collaboration on the initial Paul-Trap experiment. Thanks go to Dr. David Gregory, Dr. Egor Chasovskikh for their advice on the construction of the optical tweezer and Chun-Hao Hwu for his assistance in the lab. They gratefully acknowledge the technical support from Michel Moret.

## AUTHOR DECLARATIONS

### Conflict of Interest

The authors have no conflicts to disclose.

## DATA AVAILABILITY

The data that support the findings of this study are available from the corresponding author upon reasonable request.

# Opto-mechanical expulsion of individual micro-particles by laser-induced shockwave in air

M. C. SCHROEDER[1], U. ANDRAL[1], AND J.-P. WOLF[1,a)]

[1]*Groupe de Physique Appliquée, Université de Genève, 1211 Geneva 4, Switzerland*

a)Author to whom correspondence should be addressed: jean-pierre.wolf@unige.ch

## Supplementary

**Droplet Free Fall Experiment**

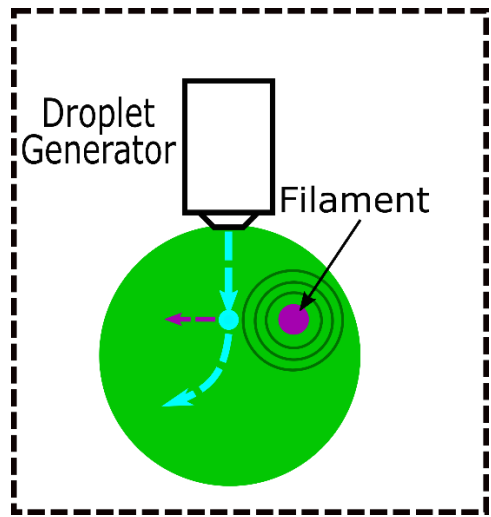

FIG. S1. Sketch of the free fall experiment from the perspective of the CCD. Droplet trajectory shown in blue. Filament position and direction of force shown in purple. Filaments were generated when the droplet was in the same plane, approximately 700 $\mu$s after droplet generation. The green disk represents the probe Nd:YAG laser beam,

The droplet free fall experiment was realized on a similar experimental layout to the tweezer experiment. The trapping cell assembly, marked in Figure 1a was replaced by a piezo-driven droplet generator (*Piezodrop*) and the focal length $L_{Fil}$ = 200 mm of the filamentation lens changed to $L_{Fil} = 150\ mm$. Figure S1 shows a sketch of the experiment from the perspective of the CCD camera. The droplet generator was connected to a digital delay generator, as was the 10 ns-pulsed

Nd:YAG laser. Time zero was given by the TiSA system, which was set to a repetition rate of 10 Hz. The trigger for the droplet generator was chosen in such a way, that the droplet would be dispensed slightly before the TiSA laser fired. The droplet could thus fall some distance from the nozzle and be perpendicular to the filament when it was generated. This maximizes the lateral motion of the droplet from the momentum transferred by the shockwave, marked by the purple arrow in figure S1. The trajectory could then be observed by utilizing a stroboscopic method. The delay for the trigger signal for the pulsed YAG was set such that we could observe the droplets position on the CCD camera at a given time $\Delta t_{YAG}$ before or after the filament was generated. By increasing the delay from zero the droplets position at different times can be measured and thus a trajectory can be plotted. Notice that the trajectory reflects then the motion of many individually dispensed droplets, and not the actual trajectory of a single droplet. Each point of the trajectory is represents the sum of the consecutive position measurement of 100 individually dispensed, falling droplets. This is an important point as each filament firing deposits charges with long lived ions, so that each of the consecutive droplets may not see the same electrostatic potential environment.

Trajectories were acquired with and without the presence of a filament, in order to account for natural drift when the droplet is ejected from the nozzle of the generator. Figures S2 and S3 show the trajectory of the droplets under the presence of a filament at different times. The filament positions are marked as circles in the plot and the separation between them and the droplets range from $480\ \mu m$ up to $1.9\ mm$. Each droplet position is derived from the mean of 100 single frames at any given time delay. Time delays span from $\Delta t = 100\ \mu s$ up to $1\ ms$ in steps of $100\ \mu s$ and then from $1\ ms$ up to $3\ ms$ in steps of $1\ ms$. The filament was generated $700\ \mu s$ after the droplet is ejected from the nozzle, at the moment when it is perpendicular to the filament.

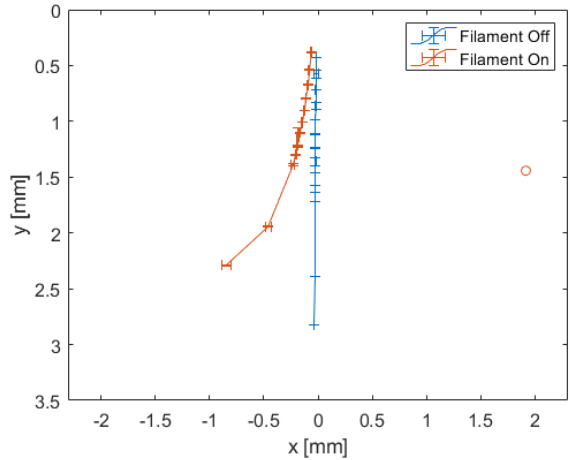

FIG. S2. Droplet trajectory reconstructed from the stroboscopic measurements done with the synchronized piezo-driven droplet generator. Femtosecond laser pulse with a peak power of approximately $I_0 = 3.8\ GW$ generated plasma filaments at the position marked with a circular point every 100 ms. Filaments were generated at $\Delta t = 700\ \mu s$, when perpendicular to the droplets.

The droplet trajectories clearly show the impact of the laser filament, as shown in Figure S2. During the recorded 3 ms duration fall, the droplet is pushed laterally by approximately 2.7 mm, as compared to the case without filament. This behavior of induced lateral motion could be observed for varying incident intensities. However, a main issue was to distinguish clearly between contributions from the shockwave and from electrostatic effects. A well-known side effect of piezo-driven droplet generators is electric charging of the droplets when they are expelled from the nozzle. Coulomb interaction between the droplet and the plasma generated during filamentation may thus occur and modify the droplet trajectories. Additionally, a repetitive generation of filaments at a repetition rate of 10 Hz causes cumulative charging of the surrounding environment and the droplet by the residual ions left after the plasma relaxation. This results in electrostatic artifacts, which prevent a quantitative measurement of the applied force due to the shockwave. These ambiguities necessitated an improvement of our ansatz and a more elaborate and precise measurement strategy.

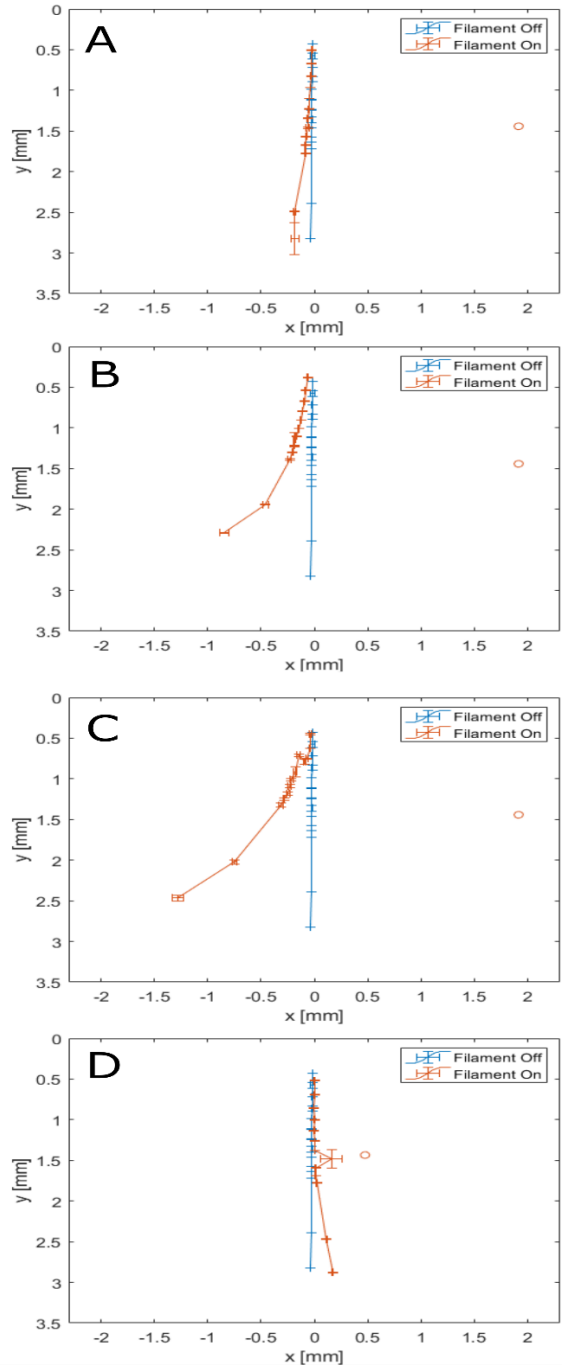

FIG. S3. Free falling droplet measurements for different incident intensities and two separation distances between the filament core and the falling water droplet. All figures include a blue reference curve, which shows the unperturbed trajectory of the falling droplets when no filament is generated. The red curves show the measured deviation caused by the presence of a single filament during the fall. All filaments were generated at the same relative time of 0.7 ms after droplet ejection. At his time the droplet was perpendicular to them along the horizontal plane. A-C) 1.2, 3.8, 10.6 GW laser pulse peak power respectively at 1.92 mm separation distance between droplet and filament core. Deviations of the droplet trajectory away from the filament position are noticeable in all cases. Motion along the x-axis increases with increased initial peak power. D ) Also using 10.6 GW peak power, but at 0.48 mm separation distance between droplet and filament core. The droplet now moves toward the filament position, indicating the presence of additional interactions apart from the shockwave.

**Pictures of the optical trap, trapped bead and trapped bead cluster**

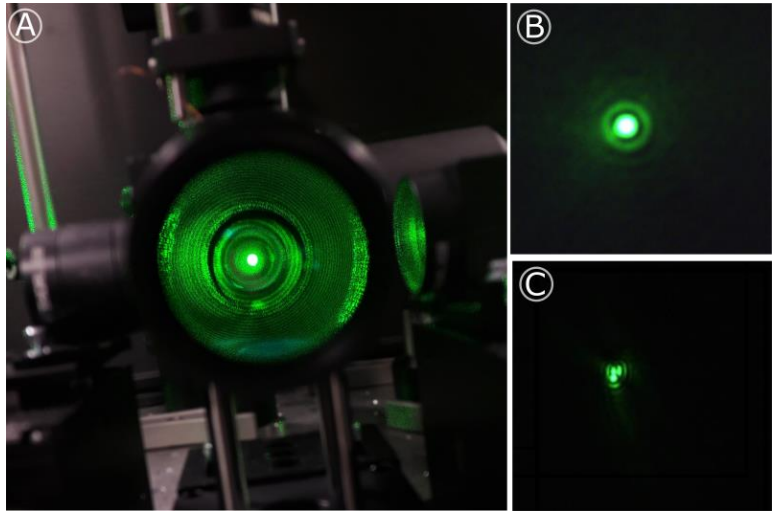

FIG. S4. Images of successfully trapped microspheres. Figure A shows a trapped sphere in the enclosed trapping cell from the perspective of the CCD camera. Figure B and C are taken using a 20x microscope objective and show two possible cases for successful trapping. In B a single microsphere is trapped. This is the usable case for the performed experiment. Figure C shows a cluster of microspheres, which is trapped. These clusters could not be used for the expulsion or displacement measurements due to the skewered trapping behavior and increased cross-section. If a cluster was observed the trap was reset and loaded again

**Estimation of the error due to a trap misalignment**

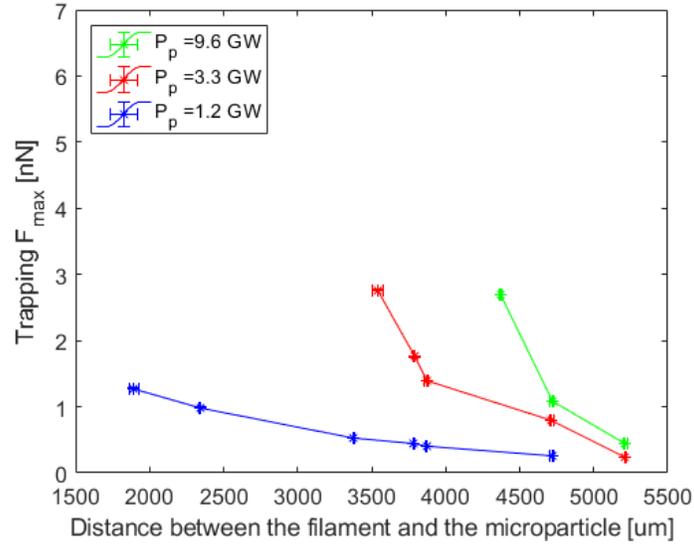

FIG. S5. The model calculations in Fig. 5 are performed with the assumption of an ideal trapping case. This means we assume perfect alignment of the trapping beams with both foci overlapping in all spatial dimensions. The sub-optimal case, where the trapping beams are misaligned, but less efficient trapping is still possible, needs to be considered as well. We cannot guarantee perfect overlap in all dimensions. As alignment is done on a pinhole with an aperture almost equal to the beam waist, misalignment in the plane orthogonal to the trapping beams is kept to a minimum with variations of ±0.5 μm. Along the beam axis overlap is more difficult to achieve and a separation of the foci of a few micrometers can occur. The separation of the foci itself is limited to approximately 9 μm, based on our calculated model. Further separation would result in the loss the trapping potential. A further separation would result in the loss of potential suitable for trapping. These calculations give us a lower bound for the radial trapping force and applied acoustic force. The results are shown above.

**Power law fitting of the distance dependence of the shock-wave force in the case of a cylindrical shock wave (Bethe et al)**

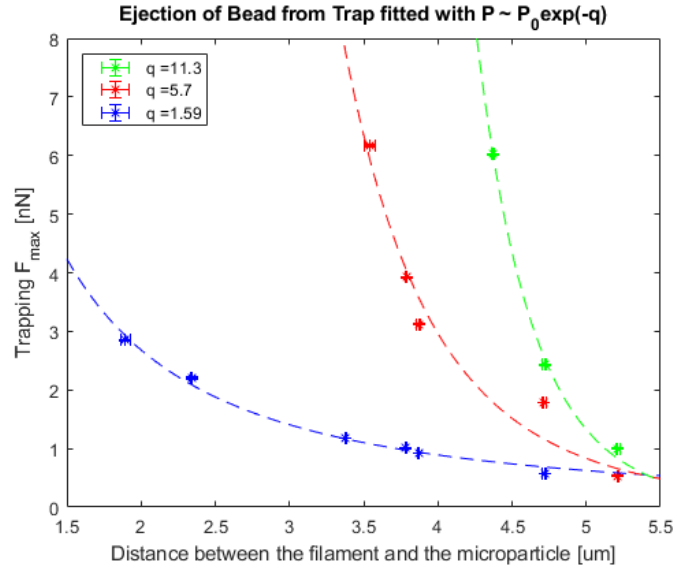

Fig. S6. Experimental data fitted with the expected pressure amplitude decay for a cylindrical shock wave as described by Bethe et al. The amplitude $\Delta P(r)$ of the pressure decrease is expected to scale with radius $r$ according to $\Delta P(r) \sim P\_0 \ r^{-q}$.